# GATED RECURRENT NETWORKS FOR SEIZURE DETECTION

*M. Golmohammadi, S. Ziyabari, V. Shah, E. Von Weltin, C. Campbell, I. Obeid and J. Picone*

Neural Engineering Data Consortium, Temple University, Philadelphia, Pennsylvania, USA
{meysam, saeedeh, vinitshah, eva.vonweltin, christopher.campbell, obeid, picone}@temple.edu

*Abstract*— Recurrent Neural Networks (RNNs) with sophisticated units that implement a gating mechanism have emerged as powerful technique for modeling sequential signals such as speech or electroencephalography (EEG). The latter is the focus on this paper. A significant big data resource, known as the TUH EEG Corpus (TUEEG), has recently become available for EEG research, creating a unique opportunity to evaluate these recurrent units on the task of seizure detection. In this study, we compare two types of recurrent units: long short-term memory units (LSTM) and gated recurrent units (GRU). These are evaluated using a state of the art hybrid architecture that integrates Convolutional Neural Networks (CNNs) with RNNs. We also investigate a variety of initialization methods and show that initialization is crucial since poorly initialized networks cannot be trained. Furthermore, we explore regularization of these convolutional gated recurrent networks to address the problem of overfitting. Our experiments revealed that convolutional LSTM networks can achieve significantly better performance than convolutional GRU networks. The convolutional LSTM architecture with proper initialization and regularization delivers 30% sensitivity at 6 false alarms per 24 hours.

## I. INTRODUCTION

Diagnosis of clinical conditions such as epilepsy are dependent on electroencephalography (EEG), the recording of the brain's electrical activity through electrodes placed on the scalp. Delivering a conclusive diagnosis of a brain-related illness without an EEG is often unfeasible [1]. The large amounts of time required by specialized neurologists to interpret these records, has created a workflow bottleneck – neurologists are overwhelmed with the amount of data that needs to be manually reviewed [2]. There is a great need for partial or complete automation of the EEG analysis process, and automated technology is slowly emerging to fill this void [3][4]. Automatic analysis of EEG scans reduces time to diagnosis, reduces error and enhances a neurologist's ability to administer medications. The ability to search EEG records symbolically greatly accelerates the review process.

In this paper, we focus specifically on the problem of seizure detection. Many algorithms have been applied to this problem including time–frequency analysis methods [5][6], nonlinear statistical models [7] and more modern machine learning approaches such as neural networks and support vector machines [8]. Despite much progress, current EEG analysis methodologies are far from perfect with many being considered impractical due to high false detection rates [9]-[11]. The machine learning challenges of this application are described extensively in [1].

A significant big data resource, known as the TUH EEG Corpus (TUEEG) [12], has become available for EEG interpretation creating a unique opportunity to advance technology. Using a subset of this data that has been manually annotated for seizure events [13], a novel deep structure has been recently introduced which achieves a low false alarm rate on EEG signals [14]. This system integrates convolutional neural networks (CNNs) with recurrent neural networks (RNNs) to deliver state of the art performance. In this paper, our goal is to investigate the use of RNNs using the high-performance architecture described in [13]. We also explore improved initialization methods and regularization approaches.

## II. RECURRENT NEURAL NETWORKS

A recurrent neural network (RNN) is an extension of a conventional feedforward neural network which can handle a variable-length input. The RNN handles the variable-length sequence by having a recurrent hidden state whose activation at each time is dependent on that of the previous time. Standard RNNs are hard to train due to the well-known vanishing or exploding gradient problems [15][16]. To address the vanishing gradient problem, the gated recurrent network architectures such as long short-term memory (LSTM) [17] unit and gated recurrent unit (GRU) were proposed [18].

LSTM was presented in [17]. The most commonly used architecture was described in [19], and is formulated as:

$$i_t = \sigma(U^i x_t + W^i s_{t-1} + p^i \circ c_{t-1} + b^i), \quad (1)$$

$$f_t = \sigma(U^f x_t + W^f s_{t-1} + p^f \circ c_{t-1} + b^f), \quad (2)$$

$$c_t = f_t \circ c_{t-1} + i_t \circ g(U^c x_t + W^c s_{t-1} + b^c), \quad (3)$$

$$o_t = \sigma(U^o x_t + W^o s_{t-1} + p^o \circ c_t + b^o), \quad (4)$$

$$s_t = o_t \circ g(c_t). \quad (5)$$

where $i_t$, $f_t$, $c_t$, $o_t$, and $s_t$ are the input gate, forget gate, cell state, output gate and block output at time instance t, respectively; $x_t$ is the input at time t; $U^*$, and $W^*$ are the weight matrices applied on input and recurrent hidden units, respectively; $\sigma(.)$ and $g(.)$ are the sigmoid and tangent activation functions, respectively; $p^*$ and $b^*$ are the peep-hole connections and biases, respectively; and ∘ means element-wise product.

As an alternative to the LSTM, the Gated Recurrent Unit (GRU) architecture was proposed in [20]. In [18], a GRU architecture was found to achieve better performance than LSTM on some tasks. A GRU is formulated as:

$$r_t = \sigma(U^r x_t + W^r s_{t-1} + b^r), \quad (6)$$
$$z_t = \sigma(U^z x_t + W^z s_{t-1} + b^z), \quad (7)$$
$$\tilde{s}_t = g(U^s x_t + r_t \circ (W^s s_{t-1}) + b^s, \quad (8)$$
$$s_t = z_t \circ s_{t-1} + (1 - z_t) \circ \tilde{s}_t. \quad (9)$$

As one can see, a GRU architecture is similar to LSTM but without a separate memory cell. Unlike LSTM, a GRU does not include output activation functions and peep-hole connections. It also integrates the input and forget gates into an update gate, $z_t$, to balance between the previous activation, $s_{t-1}$, and the candidate activation, $\tilde{s}_t$. The reset gate, $r_t$, allows it to forget the previous state [21].

### III. EXPERIMENTAL DESIGN

Our basic architecture, which employs a convolutional recurrent neural network, is presented in Figure 1. In this architecture, we integrate 2D CNNs, 1-D CNNs and LSTM networks to better exploit long-term dependencies. This structure currently uses LSTMs. However, we can easily replace LSTMs with GRUs. Feature extraction is performed using a fairly standard linear frequency cepstral coefficient-based feature extraction approach (LFCCs) popularized in applications such as speech recognition [22][23]. We also use first and second derivatives of the features since these provide a small improvement in performance [22].

Drawing on a video classification analogy, input data to the first layer of CNNs is composed of frames distributed in time where each frame is an image of width (W) equal to the length of a feature vector, the height (H) equals the number of EEG channels, and the number of image channels (N) equals one. Input data consists of T frames where T is equal to the window length multiplied by the number of samples per second.

In our optimized system with a window duration of 21 seconds, the first 2D convolutional layer filters 210 frames (T = 21 × 10) of EEGs distributed in time with a size of 26 × 22 × 1 (W=26, H=22, N=1) using 16 kernels of size 3 × 3 and with a stride of 1. The first 2D max pooling layer takes as input a vector which is 260 frames distributed in time with a size of 26 × 22 × 16 and applies a pooling size of 2 × 2. This process is repeated two times with two 2D convolutional layers with 32 and 64 kernels of size 3 × 3 respectively and two 2D max pooling layers with a pooling size 2 × 2.

The output of the third max pooling layer is flattened to 210 frames with size of 384 × 1. Then a 1D convolutional layer filters the output of the flattening layer using 16 kernels of size 3 which decreases the dimensionality in space to 210 × 16. Then we apply a 1D maxpooling layer with a size of 8 to decrease the dimensionality to 26 × 16. This is the input to a deep bidirectional LSTM network where the dimensionality of the output space is 128 and 256. The output of the last bidirectional LSTM layer is fed to a 2-way sigmoid function which produces a final classification of an epoch. Epochs are typically 1 sec in duration.

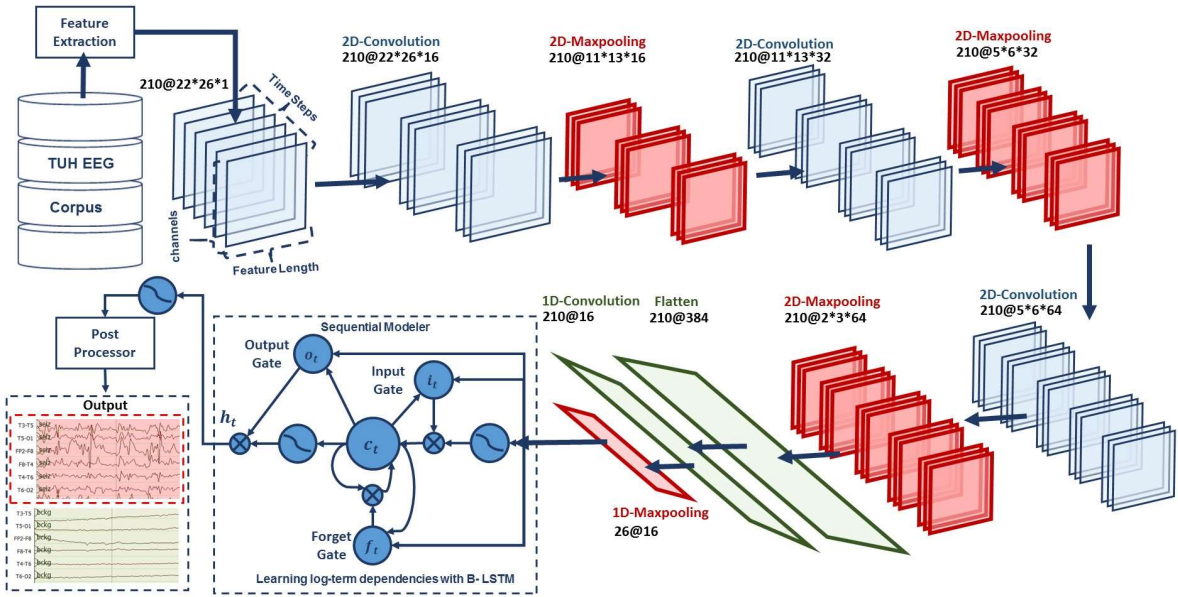

Figure 1: A deep recurrent convolutional architecture for two-dimensional decoding of EEG signals that integrates 2D CNNs, 1-D CNNs and LSTM networks is shown. In this structure, LSTMs can be easily replaced by GRUs.

To overcome the problem of overfitting and force the system to learn more robust features, regularizations are used in first two layers of CNNs. To increase non-linearity, Exponential Linear Units (ELU) are used [24]. Adam is used in the optimization process along with a mean squared error loss function [14].

## IV. RESULTS

The lack of big data resources that can be used to train sophisticated statistical models compounds a major problem in automatic seizure detection. Inter-rater agreement for this task is low, especially when considering short seizure events [1]. Manual annotation of a large amount of data by a team of certified neurologists is extremely expensive and time consuming. It is difficult to employ large numbers of board-certified neurologists to perform this task. In this study, we are reporting results on the TUH EEG Seizure Corpus (TUSZ) [13]. This dataset, which is publicly available, is a subset of the TUH EEG Corpus that focuses on the problem of seizure detection. A summary of the corpus (v1.1.1) is shown in Table 1.

A comparison of the performance of the convolutional recurrent neural networks using GRU and LSTM architectures, for sensitivity in range of 30%, are shown in Table 2 The related DET curve is illustrated in Figure 2 These systems were evaluated using a method of scoring popular in the EEG research community known as the overlap method [25]. True positives (TP) are defined as the number of epochs identified as a seizure in the reference annotations and correctly labeled as a seizure by the system. True negatives (TN) are defined as the number of epochs correctly identified as non-seizures. False positives (FP) are defined as the number of epochs incorrectly labeled as seizure while false negatives (FN) are defined as the number of epochs incorrectly labeled as non-seizure. Sensitivity shown in Table 2 is computed as TP/(TP+FN). Specificity is computed as TN/(TN+FP). The false alarm rate is the number of FPs per 24 hours.

By comparing the results of CNN/LSTM with CNN/GRU, demonstrated in Figure 2, we find that in lower false positive rates, CNN/LSTM has significantly better performance from CNN/GRU, due to the fact that while GRU unit controls the flow of information like the LSTM unit, but it does not have a memory unit. LSTMs can remember longer sequences better than GRUs and outperform it in this task, since seizure detection requires modeling long distance relationships. Additionally, the training time of CNN/GRU was 10% less than CNN/LSTM. Hence, the training time of these two systems is comparable, since most of the cycles are used for training of convolutional layers.

| Description | Train | Eval |
|---|---|---|
| Patients | 196 | 50 |
| Sessions | 456 | 230 |
| Files | 1505 | 984 |
| Seizure (secs) | 51,140 | 53,930 |
| Non-Seizure (secs) | 877,821 | 547,728 |
| Total (secs) | 928,962 | 601,659 |

Table 1. An overview of the TUH EEG Seizure Corpus (v1.1.1)

| System | Sensitivity | Specificity | FA/24 Hrs. |
|---|---|---|---|
| CNN/GRU | 30.83% | 91.49% | 21 |
| CNN/LSTM | 30.83% | 97.10% | 6 |

Table 2. Recognition results for convolutional recurrent neural networks using GRU and LSTM architectures, for sensitivity in range of 30%.

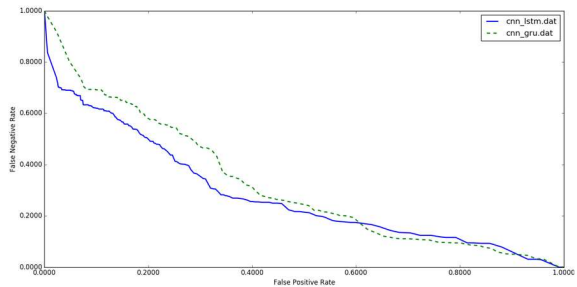

Figure 2. DET Curves for convolutional recurrent neural networks using GRU and LSTM architectures.

In neural networks, determining the proper initialization strategy for the parameters in the model is part of the difficulty in training. Hence, we investigated a variety of initialization methods using the structure introduced in Figure 1. These results are presented in Table 4. The related DET curve is illustrated in Figure 4. In our experiments, we observed that the proper initialization of weights in a convolutional recurrent neural network is critical to convergence. For example, initialization with zeros or ones methods completely stalled the convergence process. Also, as we can see in Table 4, the performance of the system for the same sensitivity of 30% can change from 7 to 40, for different initialization methods. This decrease in performance and deceleration of convergence arises because some initializations can result in the deeper layers receiving inputs with small variances, which in turn slows down back propagation, and retards the overall convergence process.

The best performance is achieved using orthogonal initialization. This method is a simple yet effective way of combatting exploding and vanishing gradients. Orthogonal matrices preserve the norm of a vector and their eigenvalues have absolute value of one. This means that, no matter how many times we perform repeated matrix multiplication, the resulting matrix doesn't explode or vanish [26]. Also in orthogonal matrices, columns and rows are all orthonormal to one another, which helps the weights to learn different input features.

| Initialization | Sensitivity | Specificity | FAs |
|---|---|---|---|
| Orthogonal [26] | 30.8% | 96.9% | 7 |
| Lecun Uniform [27] | 30.3% | 96.5% | 8 |
| Glorot Uniform [28] | 31.0% | 94.2% | 13 |
| Glorot Normal [28] | 29.5% | 92.4% | 18 |
| Variance Scaling [26] | 31.8% | 92.1% | 19 |
| Lecun Normal [27] | 31.8% | 92.1% | 19 |
| He Normal [29] | 31.3% | 91.1% | 22 |
| Random Uniform [26] | 30.2% | 90.0% | 25 |
| Truncated Normal [26] | 31.6% | 87.8% | 31 |
| He Uniform [29] | 29.2% | 85.1% | 40 |

Table 4. Results for CNN/LSTM, for sensitivity in the range of 30%, using different initialization methods.

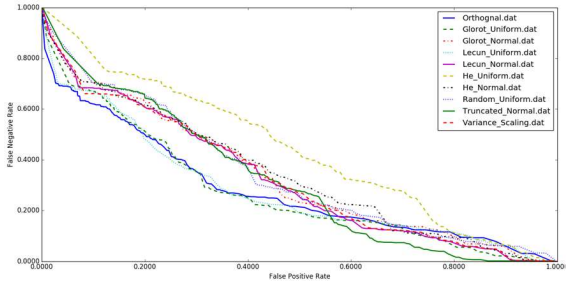

Figure 4. DET curves for the CNN/LSTM architecture using different initialization methods.

Overfitting is a serious problem in deep neural nets with many parameters. In this study, we used five popular regularization methods to address this problem. By using L1, L2 and L1/L2 techniques, we apply penalties on layer parameters during optimization [30]. These penalties are incorporated in the loss function that the network optimizes. In an alternative approach, we used dropout to prevents units from co-adapting too much by randomly dropping units and their connections from the neural network during training [30]. Also, we studied the impact of introducing zero-centered Gaussian noise to the network [30]. The results of these experiments are presented in Table 3 along with a DET curve in Figure 3.

While generally L1/L2 has the best performance, as we move towards a low FA rate, dropout delivers a lower FA rate. Additionally, we found that the primary error modalities observed were false alarms generated during brief delta range slowing patterns such as intermittent rhythmic delta activity. Our closed-loop experiments showed us that all the regularizing methods presented in Table 3 are playing an important role in increasing the false alarms of slowing patterns. Even though dropout is effective in CNNs, when dropout is placed over kernels it leads to diminished results. To solve this problem, in our future work, an efficient Bayesian convolutional neural network is being explored that places a probability distribution over the CNN's kernels [31]. This approach offers better robustness to overfitting on small data and show improve the robustness of our training process.

| Initialization | Sensitivity | Specificity | FAs |
|---|---|---|---|
| L1/L2 | 30.8% | 97.1% | 6 |
| Dropout | 30.8% | 96.9% | 7 |
| Gaussian | 30.8% | 95.8% | 9 |
| L2 | 30.2% | 95.6% | 10 |
| L1 | 30.0% | 43.7% | 276 |

Table 3. Recognition results for convolutional LSTM architecture, for sensitivity in range of 30%, using different regularizations.

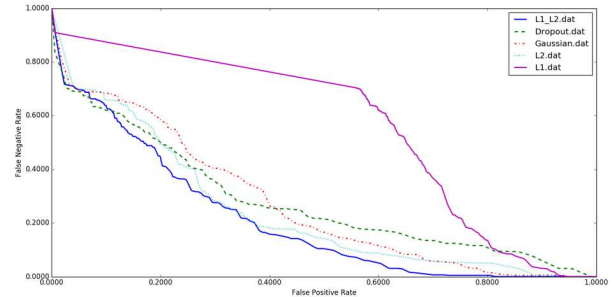

Figure 3. DET curves for the CNN/LSTM architecture using different regularizations.

SUMMARY

In this paper, we investigated two deep learning architectures (LSTM and GRU) for automatic classification of EEGs using CNNs. LSTMs outperformed GRUs. We also studied initialization and regularizations of these networks. In future research, we are designing a more powerful architecture based on reinforcement learning concepts. We are also optimizing regularization and initialization algorithms for these approaches. Our goal is to approach human performance which is in the range of 75% sensitivity with a false alarm rate of 1 per 24 hours [11]. Robust training procedures are needed to make this technology relevant to a wide range of healthcare applications.

ACKNOWLEDGEMENTS

Research reported in this publication was most recently supported by the National Human Genome Research Institute of the National Institutes of Health under award number U01HG008468. The content is solely the responsibility of the authors and does not necessarily represent the official views of the National Institutes of Health. This material is also based in part upon work supported by the National Science Foundation under Grant No. IIP-1622765. Any opinions, findings, and conclusions or recommendations expressed in this material are those of the author(s) and do not necessarily reflect the views of the National Science Foundation. The TUH EEG Corpus work was funded by (1) the Defense Advanced Research Projects Agency (DARPA) MTO under the auspices of Dr. Doug Weber through the Contract No. D13AP00065, (2) Temple University's College of Engineering and (3) Temple University's Office of the Senior Vice-Provost for Research.